\documentclass[12pt,aps,onecolumn,superscriptaddress]{revtex4}

\usepackage{graphicx}
\usepackage{amsmath}
\usepackage{hyperref}

\usepackage{ifthen}
\newboolean{linenum}

\begin{document}

\title{\boldmath {Cross section of the process 
$e^+e^-\to n\bar{n}$ near the threshold} }
\ifthenelse{\boolean{linenum}}{
\linenumbers}{}

\newcommand{\binp}{\affiliation{Budker Institute of Nuclear Physics,
SB RAS, Novosibirsk 630090, Russia}}
\newcommand{\nsu}{\affiliation{Novosibirsk State University,
Novosibirsk 630090, Russia}}

\author{M.~N.~Achasov} \binp\nsu
\author{A.~Yu.~Barnyakov} \binp\nsu
\author{E.~V.~Bedarev} \binp\nsu\
\author{K.~I.~Beloborodov} \binp\nsu
\author{A.~V.~Berdyugin} \binp\nsu
\author{A.~G.~Bogdanchikov}\binp
\author{A.~A.~Botov}\binp
\author{T.~V.~Dimova}\binp\nsu
\author{V.~P.~Druzhinin}\binp\nsu
\author{V.~N.~Zhabin}\binp\nsu
\author{Yu.~M.~Zharinov}\binp
\author{L.~V.~Kardapoltsev}\binp\nsu
\author{A.~S.~Kasaev}\binp
\author{A.~A.~Kattsin}\binp
\author{D.~P.~Kovrizhin} \binp
\author{A.~A.~Korol}\binp\nsu
\author{A.~S.~Kupich} \binp\nsu
\author{A.~P.~Kryukov} \binp
\author{A.~P.~Lysenko} \binp
\author{N.~A.~Melnikova} \binp
\author{N.~Yu.~Muchnoi} \binp\nsu
\author{A.~E.~Obrazovsky} \binp
\author{E.~V.~Pakhtusova} \binp
\author{K.~V.~Pugachev} \binp\nsu
\author{S.~A.~Rastigeev} \binp
\author{Yu.~A.~Rogovsky} \binp\nsu
\author{A.~I.~Senchenko} \binp
\author{S.~I.~Serednyakov} \email{S.I.Serednykov@inp.nsk.su} \binp\nsu
\author{Z.~K.~Silagadze} \binp\nsu
\author{I.~K.~Surin} \binp
\author{Yu.~V.~Usov} \binp
\author{A.~G.~Kharlamov}\binp\nsu
\author{D.~E.~Chistyakov} \binp\nsu
\author{Yu.~M.~Shatunov} \binp
\author{S.~P.~Sherstyuk} \binp\nsu
\author{D.~A.~Shtol} \binp

\date{ }

\begin{abstract}
The $e^+e^-\to n\bar{n}$ cross section was measured at 
center of mass (c.m.) energies 
from the threshold to 1908 MeV. 
The experiment to measure the cross section  has been
carried out at the VEPP-2000 $e^+e^-$ collider in 13 energy points. 
The  SND detector is used to detect the produced 
neutron-antineutrons  ($n\bar{n}$) events. A special time measurement
system on the calorimeter was used to select the time-delayed $n\bar{n}$
events. The measured $e^+e^-\to n\bar{n}$ cross section is 0.4--0.6 nb.  
The neutron effective timelike form factor  in the energy range
under study varies from 0.3 to 0.6.
\end{abstract}
\maketitle

\section*{Introduction\label{sec:intro}}
The 
$e^+e^-$ annihilation to neutron-antineutron pairs depends on two 
form factors -  electric $G_E$ and magnetic $G_M$ :
\begin{eqnarray}
\frac{d\sigma}{d\Omega}&=&\frac{\alpha^{2}\beta}{4s}
\bigg[ |G_M(s)|^{2}(1+\cos^2\theta)\nonumber\\
&+&\frac{1}{\gamma^2}|G_E(s)|^{2}\sin^2\theta
\bigg],
\label{eqB1}
\end{eqnarray}
where $\alpha$ is the fine structure constant, 
$s=4E_b^2=E^2$, where $E_b$ is the beam  energy and $E$ is the
center-of-mass (c.m.)  energy,  $\beta = \sqrt{1-4m_n^2/s}$, $\gamma
= E_b/m_n$,   $m_n$ is the neutron
mass   and $\theta$ is the antineutron production polar angle.
The total cross section has the following form:
\begin{equation}
\sigma(s) =
\frac{4\pi\alpha^{2}\beta}{3s}(1+\frac{1}{2\gamma^2})|F(s)|^2,
\label{eqB2}
\end{equation}
where   the effective form factor  $F(s)$ is introduced:  
\begin{equation}
|F(s)|^2=\frac{2\gamma^2|G_M(s)|^2+|G_E(s)|^2}{2\gamma^2 +1 }.
\label{eqB3}
\end{equation}
The $|G_E/G_M|$ ratio can be extracted
from the analysis of the measured $\cos\theta$ distribution in 
 Eq.~(\ref{eqB1}). At the threshold  $|G_{E}| = |G_{M}|$.

  The $e^+e^-\to n\bar{n}$ process in the threshold region 
was first observed in the FENICE ~\cite{FENICE} and 
DM2  ~\cite{DM2}   experiments. 
Much more accurate measurements are being carried out  at
the VEPP-2000 e+e- collider with the SND detector 
~\cite{Art1719}, \cite{Art2021}.
At the energy above 2 GeV new data have
been obtained by the BESIII ~\cite{BES}. 
In this work we present the results of measurements at the beam energy
range $\sim$ 
15 MeV above  the nucleon threshold with a scanning step close 
to the beam energy spread.

\begin{figure*}
\centering
\includegraphics [width = 0.7\textwidth]{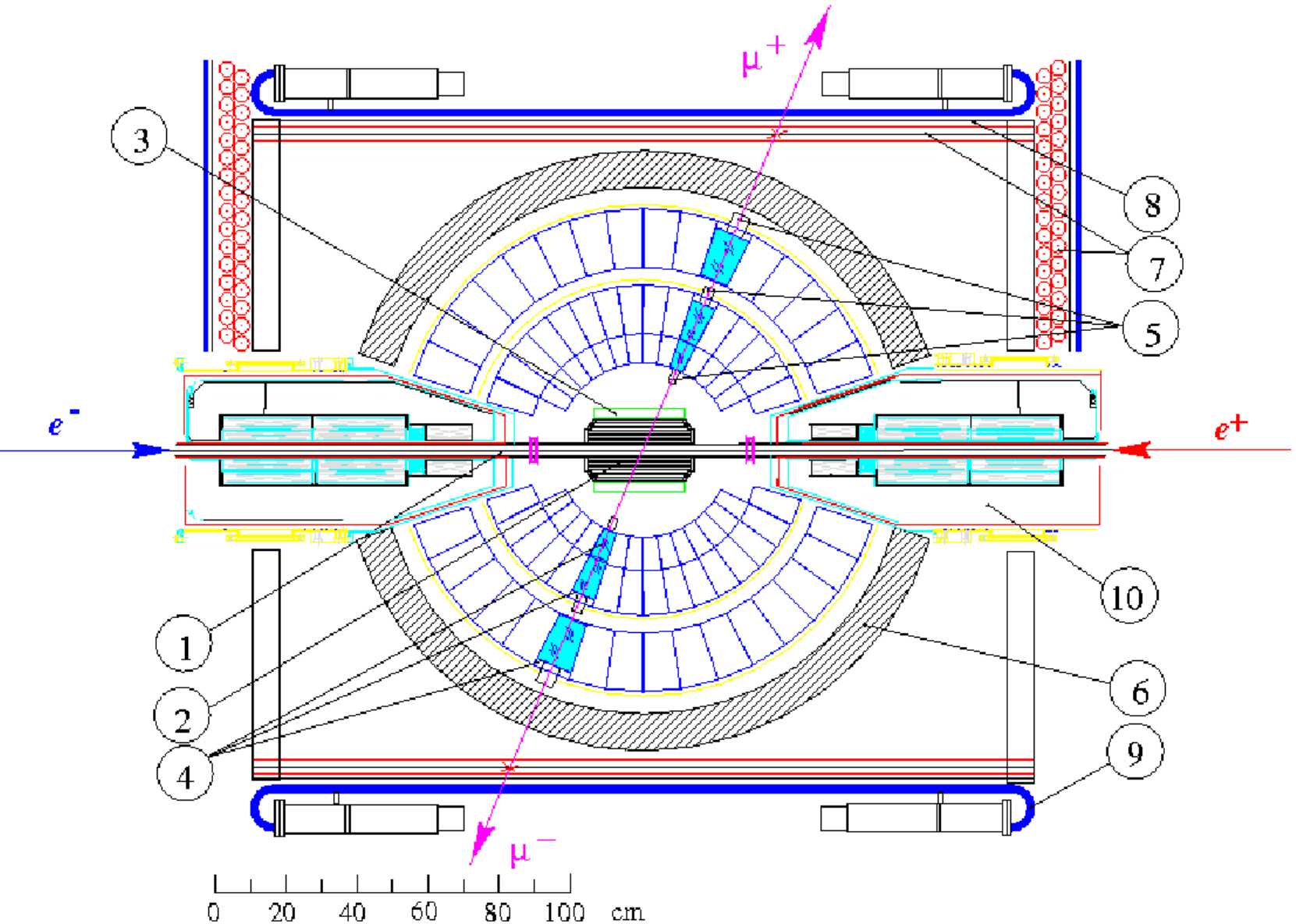}
\caption{SND detector, section along the beams: (1) beam pipe,
(2) tracking system, (3) aerogel Cherenkov counters, (4) NaI (Tl)
crystals, (5) vacuum phototriodes, (6) iron absorber, (7) proportional
tubes,
(8) iron absorber, (9) scintillation counters, (10) VEPP-2000 focusing
solenoids.}
\label{fig:sndt}
\end{figure*}

\section{Collider, detector, experiment\label{sec:Exper}}
VEPP-2000 is the $e^+e^-$ collider~\cite{VEPP2k} operating 
in the c.m. energy range 
from the hadron threshold ($E$=280 MeV) up to 2 GeV. The collider
luminosity above the nucleon threshold is of order of $5\times
10^{31}$~cm$^{-2}$s$^{-1}$. There are two
collider detectors at VEPP-2000: SND and CMD-3.

    SND (Spherical Neutral Detector) ~\cite{SNDet1}  is a non-magnetic detector,
including a  tracking system, a spherical NaI(Tl) electromagnetic calorimeter
(EMC) and a muon detector (Fig.\ref{fig:sndt}). 
The EMC is the main part of the SND used in the $n\bar{n}$ analysis. 
The thickness of EMC is 34.7 cm (13.4 radiation length).
The antineutron annihilation length in NaI(Tl)   
varies with energy from several cm close to the $n\bar{n}$ threshold 
to $\sim$15 cm at  the maximum energy ~\cite{Annih}, 
so nearly all produced antineutrons  are absorbed in the detector. 

   In  $n\bar{n}$ analysis  the EMC  is used to measure the event arrival time  
relative to the moment of beam collision. For this  on each of
the 1640 EMC counters  a  special flash ADC module
~\cite{Timr}, measuring the signal waveform, is installed. 
When fitting the flash ADC output waveform with a pre-set function, 
the time and amplitude  of the signal in the counters with more than 5
MeV energy, are calculated. 
The event time is calculated as the EMC counters energy weighted average
time ~\cite{Nimj}. 
The time resolution obtained with data  $e^+e^-\to \gamma\gamma$
events is about 0.8 ns. 
This value is greater than   Monte Carlo
resolution of 0.3 ns due to the finite time resolution of the timing
system.  So we convolve the MC $\gamma\gamma$  
time spectrum with a Gaussian  with $\sigma_{\gamma\gamma}=$0.8 ns. 
For  $e^+e^-\to n\bar{n}$
events the covolution is done with $\sigma_{nn}=1.5$--2 ns depending on the
energy.

This article presents the $n\bar{n}$ analysis results of data with the integrated
luminosity of about $100$ pb$^{-1}$,  collected in the c.m. energy
range from the threshold to $1.908$  GeV in 13 energy points.  
The value of the lowest beam energy (939.59 MeV) is just at
the process threshold.  That's why here due to the beam energy spread  
only in half of $e^+e^-$ collisions the  $n\bar{n}$ pairs are produced. 
Taking this into account the average beam energy at this point increases
to the  value of 939.93 MeV, which is above the threshold by 0.36 MeV.
The effective luminosity at this point will be two times lower than
the nominal.
The energy values and luminosity at all points are given 
in the Table \ref{tab:crsect}. 
Our measurement is the first so close to the   $n\bar{n}$ threshold.

\section{Event selection \label{sec:EvSelect}}
Antineutron from the $n\bar{n}$ pair in most cases annihilates,
producing pions, nucleons, photons and other particles, 
which deposit up to 2 GeV in EMC. The neutron from the $n\bar{n}$ pair
release a small signal in EMC, which is poorly visible against the 
background of a strong $\bar{n}$ annihilation signal, so it is not 
taken into account. The $n\bar{n}$ events are reconsructed as
multiphoton events. The antineutron production angle is defined by the
direction of the total event momentum $P=\Sigma_i E_i n_i$,    
where the summation is carried out over the calorimeter crystals, 
$E_i$ is the crystal energy, $n_i$ is the unit vector. Projections of
the vector $P$ onto the direction along and across the beams define
the polar and azimuthal antineutron angles. Some of the $n\bar{n}$  events 
are accompanied by non-beam tracks, which arise during antineutron
annihilation in the detector material.   

   Main features of   $n\bar{n}$ events, in contrast to ordinary
$e^+e^-$ annihilation events, are absence of charged tracks and photons 
from the collision region and a strong imbalance in the event momentum. 
Another feature is the presence of a significant cosmic background
events, having a signature similar to $n\bar{n}$ events and 
comparable to or exceeding the intensity of the $n\bar{n}$ events.
There is also the problem of a large background from beams of
electrons and positrons in the coillider. 
   Based on these specific features of the $e^+e^-\to n\bar{n}$ process,
we have developed the following selection conditions:

  1 - no charged tracks from the interaction region is found in an
  event (nch=0), 

  2 - the event momentum  must be significantly unbalanced  
    ($P > 0.4 E_{EMC}$), what suppress the $e^+e^-$ annihilation
    background, 

  3 - the transverse EMC energy profile of the found most energetic photon in
  an event should be broader than that expected for the electromagnetic
  shower : $L_{\gamma}>-2.5$, where $L_{\gamma}$ is the
  logarithmic parameter, describing the width of the transverse energy
  profile in EMC  ~\cite{xi2gam}, 

  4 - the veto of external muon detector is required, 
  
  5 - the events with a  cosmic-ray track in EMC, defined as a chain of crystals
  along a straight line, are rejected,  

  6 - an additional suppression of cosmic background, that have passed
  through the muon veto, is carried out by a special parameter (shcosm), that
  approximates the shape of energy deposition in the calorimeter in
  the form of ellipsoid  ~\cite{Art1719}. Further, a condition on this parameter 
  is set (shcosm$>$0.4), which leads to the suppression of cosmic background.   
  Basically, these are cosmic showers in EMC,
  
  7 - the EMC total energy deposition cut is taken to be  $E_{EMC} >
  E_b$, where $E_b$ is the beam energy. Such a cut reduces the $n\bar{n}$ 
  detection efficiency by $\sim$20\%, but almost completely removes the
  beam bqackground. 
  
    In general, the conditions  listed above   are similar to those used in our  
previous works on the $e^+e^-\to n\bar{n}$ analyses
~\cite{Art1719}, \cite{Art2021}.
After imposing the described selection conditions, we have about 400
events/pb$^{-1}$  left for further analysis.

\begin{figure*}
\includegraphics[width=0.48\textwidth]{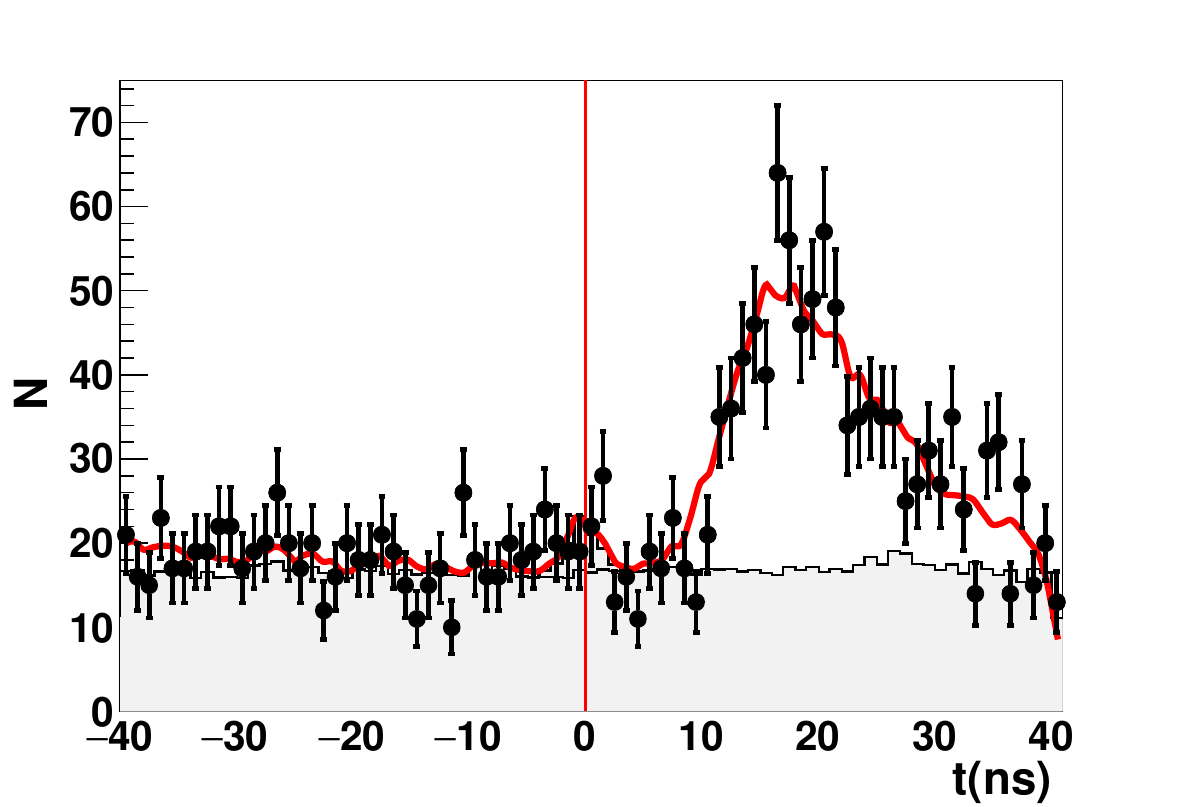} \hfill
\includegraphics[width=0.48\textwidth]{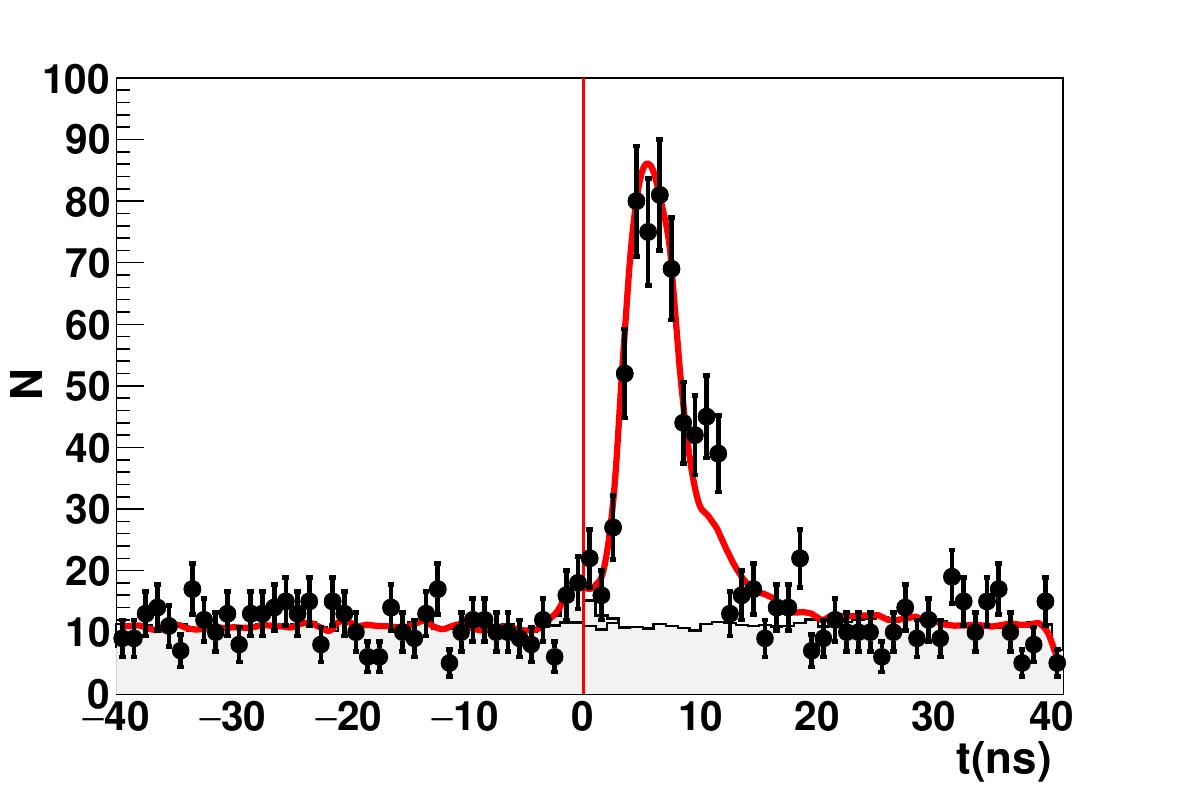}
\parbox[h]{0.48\textwidth}{\caption { 
The time distribution of  selected data events at $E_b=941$ MeV. 
The solid line (red) is the result of the fit described in the text.
Vertical line at t=0 shows the position of the beam background.}
\label{fig:hstim1}} \hfill
\parbox[h]{0.48\textwidth}{\caption {The  time distribution of  selected 
data events at $E_b=952$ MeV. The solid line (red) is the result of the fit described 
in the text. Vertical line at t=0 shows the position of the beam
background.}
\label{fig:hstim2}}
\end{figure*}
\section{Analysis of  measured time spectra 
\label{sec:Tim19}}
 Typical time spectra of events selected for analysis are shown in
Figs .~\ref{fig:hstim1}, \ref{fig:hstim2}. 
Three main components are distinguished in the figures shown: 
a beam and physical  background at t=0, a cosmic background uniform in time,  
and to the right a delayed  signal from $n\bar{n}$ events, wide in time.
The wide shape of the $n\bar{n}$ time  spectrum is explained by the spread of the
antineutron annihilation point - from the collider vacuum chamber wall
at 2 cm from the beam collision area to 70 cm on the back
wall of the calorimeter.
Respectively, the measured time spectra are fitted by the sum of these three 
contributionss in the following form :
\begin{equation}
F(t)=N_{n\bar{n}}H_{n\bar{n}}(t)+N_{\rm csm}H_{\rm csm}(t)+N_{\rm
bkg}H_{\rm bkg}(t),
\label{eq17}
\end{equation}
where  $H_{n\bar{n}}$, $H_{\rm csm}$ and $H_{\rm bkg}$ are 
normalized histograms, describing  time spectra for the $n\bar{n}$ signal,
cosmic  and beam + physical  background, respectively.
$N_{n\bar{n}}$, $N_{\rm csm}$, and $N_{\rm bkg}$ are the corresponding  event
numbers, obtained from subsequent fit. The shape of the beam+physical
background time spectrum  
$H_{\rm bkg}$ is measured at the energies below the $n\bar{n}$ threshold. 
The cosmic  time spectrum $H_{\rm csm}$ is measured with the lower EMC
threshold $0.9\cdot E_b$  in coincidence with the muon system signal.
The shape of the $H_{n\bar{n}}$ spectrum is calculated by the MC simulation the
$e^+e^-\to n\bar{n}$ process. 
But it turned out that the data $n\bar{n}$ time  
spectrum is insatisfactorily described by MC shape $H_{n\bar{n}}$.
The study of the $n\bar{n}$ simulation showed that
it contains two types of the first interaction of the antineutron with
matter: 1 - antineutron annihilation and 2 - first antineutron
scattering and then annihilation. These types have different time
spectra, and in the second case the time spectrum is wider and more
delayed. Therefore, in Eq.\ref{eq17}
the time spectrum  $H_{n\bar{n}}$ was represented
as the sum of these two contributions, and the fraction of the
scattering contibution was a free fit parameter. As a result of
fitting Eq.\ref{eq17}  with modified $H_{n\bar{n}}$, the agreement between the
data and MC in the time shape has improved significantly.
The contribution  of the first annihilation  turned out to be greater 
in data than in Monte Carlo by 1.5-2 times. The fitting curves in 
Figs .~\ref{fig:hstim1}, \ref{fig:hstim2} are modified in this way.

 The detection  cross section  of the beam+physical background  
$\sigma_{bg}=N_{bg}/L$,
where $L$ is integrated luminosity in a given energy, 
obtained during fitting, is about 5 pb and does not significantly
depend on the beam energy. The main contribution into 
physical part of $\sigma_{bg}$ comes from
the processes with neutral kaons in the final state: $e^+e^-\to
K_SK_L\pi^0$, $K_SK_L\eta$ and similar other. The role of $K_L$'s 
as a source of background  is
important for two reasons: 1 - some of $K_L$'s pass through the
calorimeter without interaction, thereby creating a momentum unbalance
similar to that in events with antineutrons; 2 - in our energy $2E_b <
2$ GeV the speed of $K_L$ can be  considerably lower than speed of light,
therefore  the signal from them is delayed as for antineutrons.    
The measured residual
cosmic background  rate has the  intensity $\sim$0.01 Hz, which
corresponds to the suppression of the number of cosmic events, that
have pass the hardware selection in the detector electronics,
approximately by $2\times 10^4$ times.

\begin{table*}
\centering
\caption{The beam energy ($E_b$), integrated luminosity ($L$),
number of selected $n\bar{n}$ events ($N_{n\bar{n}}$), the factor taking into
account radiative corrections and energy spread ($1+\delta$), corrected detection
efficiency ($\varepsilon$), measured $e^+e^-\to n\bar{n}$ cross
section $\sigma$, and neutron effective form factor ($F_n$). 
The quoted errors for $N$, $\sigma$ are statistical and 
systematic. For the detection efficiency, the systematic uncertainty
is quoted. For $F_n$, the combined statistical and systematic uncertainty is
listed.  
\label{tab:crsect}}
\begin{tabular}{cccccccc}
N & $E_b$(MeV) & $L$(pb)& $N_{n\bar{n}}$ & $1+\delta$ & $ \varepsilon$
 & $\sigma$(nb) & $F_n$\\
\hline
1 & 939.9  & 5.92 & $250\pm72$ ~& 0.602 & $0.200 \pm 0.105$ &
$ 0.352\pm0.101\pm0.193$ &$0.562\pm0.174$ \\
2 & 940.3  & 9.14 & $313\pm54$ & 0.633 & $0.121 \pm0.030$ &
$ 0.447\pm0.076\pm0.107$ &$0.555\pm0.081$ \\
3 & 941.0 & 9.70 & $597\pm41$ & 0.671 & $0.209\pm0.023$ &
$ 0.440\pm0.030\pm0.049$ &$0.467\pm0.031$ \\
4 & 942.0  & 10.12 & $680\pm37$ & 0.702 & $0.240\pm0.020$ &
$ 0.399\pm0.023\pm0.034$ &$0.390\pm0.020$ \\
5 & 943.5  & 9.81 & $747\pm36$ & 0.731 & $0.216\pm0.015$ &
$ 0.483\pm0.023\pm0.033$ &$0.382\pm0.016$ \\
6 & 945.0  & 11.45 & $921\pm38$ & 0.751 & $0.233\pm0.022$ &
$ 0.461\pm0.020\pm0.044$ &$0.345\pm0.018$ \\
7 & 947.5 & 10.41 & $947\pm37$ & 0.776 & $0.219\pm0.011$ &
$ 0.536\pm0.022\pm0.027$ &$0.340\pm0.011$ \\
8 & 948.75  & 6.42 & $611\pm 30$ & 0.787 & $0.213\pm0.014$ &
$ 0.567\pm0.028\pm0.037$ &$0.337\pm0.014$ \\
9 & 950.0  & 5.20 & $514\pm 29$ & 0.797 & $0.211\pm0.016$ &
$ 0.585\pm0.033\pm0.044$ &$0.333\pm0.016$ \\
10 & 951.0  & 5.55 & $522\pm 29$ & 0.804 & $0.223\pm0.016$ &
$ 0.525\pm0.029\pm0.039$ &$0.309\pm0.015$ \\
11 & 952.0  & 5.26 & $485\pm 26$ & 0.811 & $0.224\pm0.015$ &
$ 0.507\pm0.027\pm0.035$ &$0.298\pm0.013$ \\
12 & 953.0  & 5.68 & $513\pm 28$ & 0.818 & $0.246\pm0.017$ &
$ 0.450\pm0.024\pm0.032$ &$0.276\pm0.012$ \\
13 & 954.0  & 5.17 & $540\pm 27$ & 0.824 & $0.234\pm0.015$ &
$ 0.519\pm0.027\pm0.034$ &$0.291\pm0.012$ \\
\hline
\end{tabular}
\end{table*}

    The numbers $N_{n\bar{n}}$ of  found  events are  listed in the  
Table~\ref{tab:crsect} with the total number close to 8000.
The Table shows only statistical errors of the fitting in
$N_{n\bar{n}}$.  
A sources of systematic error  in the $N_{n\bar{n}}$ 
number can be uncertainties in the magnitide and shape of the 
time spectrum of the beam and cosmic background. The error introduced
by these sources is about 10 events what is  much lower than
statitistical errors in the Table~\ref{tab:crsect} and is  not taken
into account in what follows. 

\begin{figure*}
\includegraphics[width=0.47\textwidth]{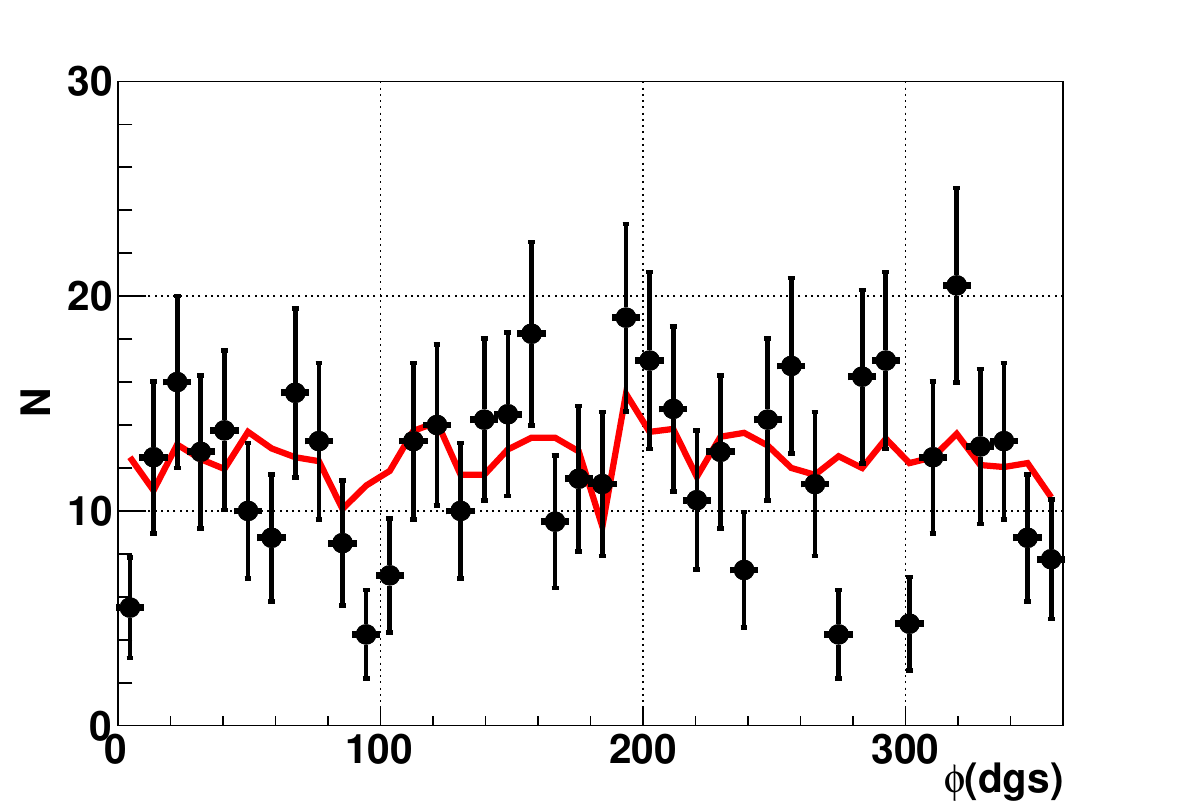} \hfill
\includegraphics[width=0.47\textwidth]{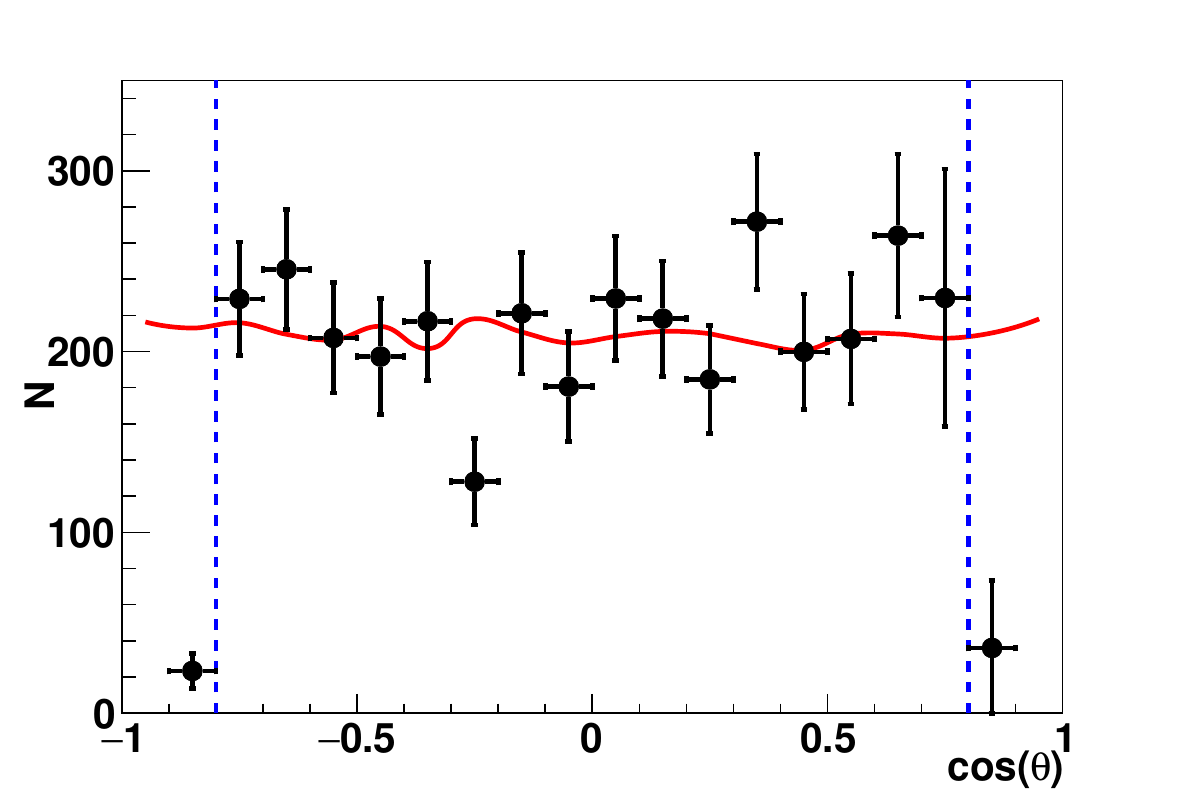} \\
\parbox[h]{0.47\textwidth}{\caption {The antineutron azimuthal angle 
distribution  for data (points with error bars) and MC (horizontal line) at
$E_b=953$ MeV. }
\label{fig:phinn}} \hfill
\parbox[h]{0.47\textwidth}{\caption {The antineutron polar angle  distribution 
for data (points with error bars) and MC (horizontal line) at  
$E_b=940$ MeV. Dotted vertical lines correspond to the polar angle cutoff. }
\label{fig:costh}}
\end{figure*}
%
\begin{figure*}
\includegraphics[width=0.48\textwidth]{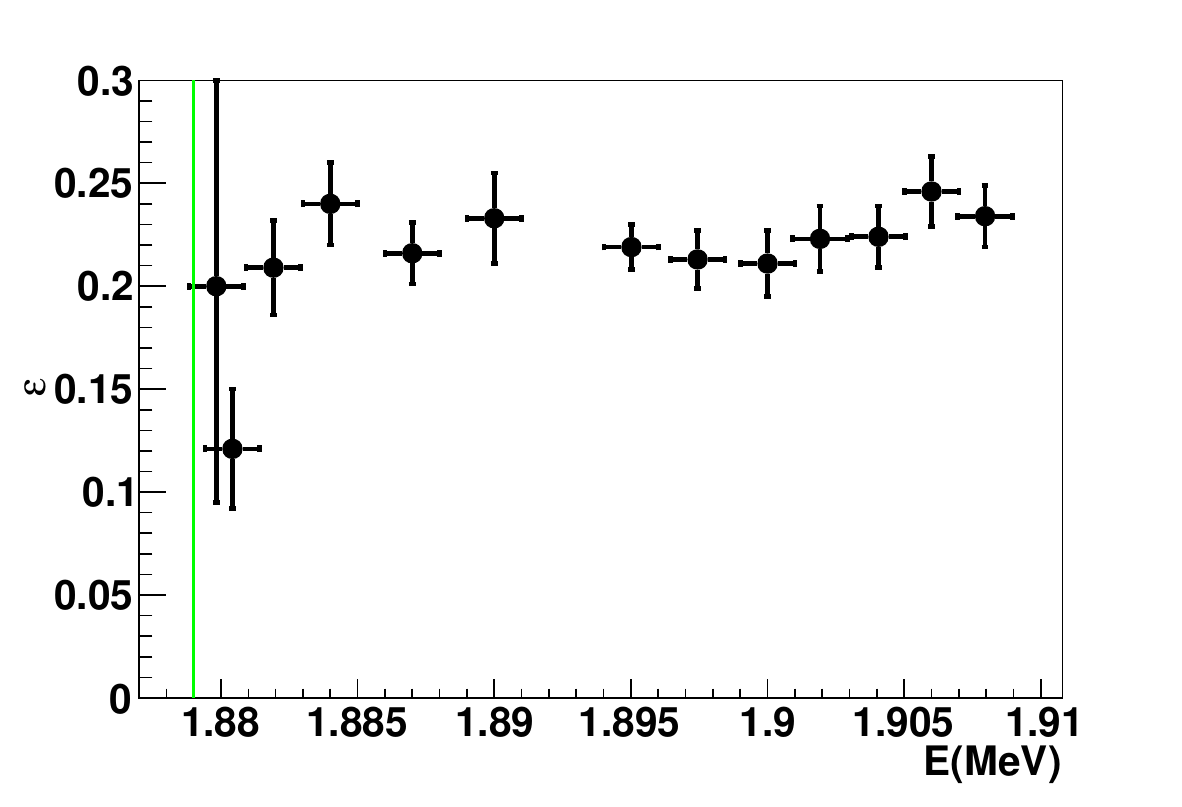} \hfill
\includegraphics[width=0.48\textwidth]{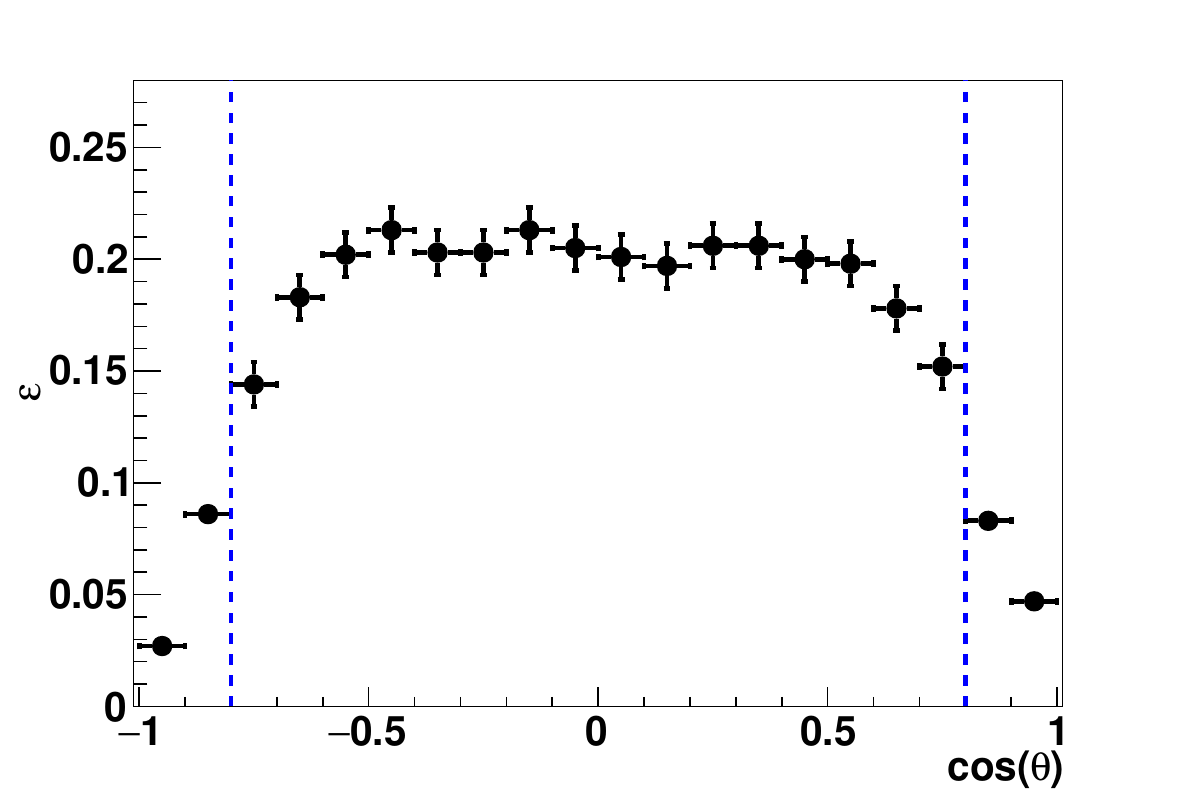} \\ 
\parbox[h]{0.47\textwidth}{\caption { The corrected MC detection efficiency  
versus energy. Vertical line corresponds to $n\bar{n}$ threshold. }
\label{fig:efenr}} \hfill
\parbox[h]{0.47\textwidth}{\caption{The MC detection efficiency
as a function of antineutron $\cos{\theta}$ at $E_b$=940  MeV.
Dotted vertical lines correspond to the polar angle cutoff.  
}
\label{fig:efth}}
\end{figure*}
\section{Angular  distribution\label{sec:cosinth}}
  The antineutron production angles $\theta_n$ and $\phi_n$, 
as it mentioned in Section \ref{sec:EvSelect},  are determined by the 
direction of the event momentum. The accuracy of the
antineutron angle measurement was determined by MC simulation  by comparing 
the true angle and the angle calculated  from  the direction of the
event momentum. The angular accuracy turned out to be about 6 degrees, which is
somewhat less than the angular size of one calorimeter crystal (9
degrees). 

  The distribution of the  selected data and MC simulated $n\bar{n}$
events over azimuthal angle $\phi_n$ (Fig.~\ref{fig:phinn})  is uniform as 
expected, what confirms the correct subtraction of the beam and
cosmic backgrounds, which are not uniform over $\phi_n$ angle.
Distribution over  $\cos\theta_n$ for data and MC events is shown in 
Fig.~\ref{fig:costh}. The MC simulation was done using Eq.~(\ref{eqB1})
with the assumption $|G_E|=|G_M|$. 
Since we are working in the immediate vicinity of the threshold, 
the condition $|G_E|\simeq|G_M|$ is required. In our case of spherical 
shape of the SND calorimeter,  the data distribution 
over  $\cos\theta_n$ is expected to be close to uniform, and this is
confirmed in Fig.~\ref{fig:costh}.


\section{Detection efficiency\label{sec:Efficy}}
   The $e^+e^-\to n\bar{n}$ process detection efficiency $\varepsilon$ 
in different energy points   under accepted  
selection conditions (Section \ref{sec:EvSelect}) is shown in
Fig.~\ref{fig:efenr}. When simulating the detector response  we used the MC
GEANT4 toolkit~\cite{GEANT4}, version 10.5. The angular distribution
of produced $n\bar{n}$ pairs corresponded to $|G_E|/|G_M|$=1,  
The simulation 
included the beam  energy spread $\sim~0.7$ MeV and the emission of 
photons by initial electrons and positrons. The simulation also took
into account non-operating detector channels as well as overlaps of
the beam background with recorded events. To do this, during the
experiment, with a pulse generator, synchronized with the moment
of beam collision, special superposition events were recorded, which
were subsequently superimposed on MC events.  The detection efficiency 
$\varepsilon$ in Fig.~\ref{fig:efenr} is corrected for the difference 
between the data and MC. This correction is discussed later.
Numerical values of the  efficiency are given in the  Table~\ref{tab:crsect}.  
~In Fig.~\ref{fig:efth} the detection efficiency 
versus  $\cos\theta$ of antineutron  is shown. 

\begin{figure*}
\includegraphics[width=0.48\textwidth]{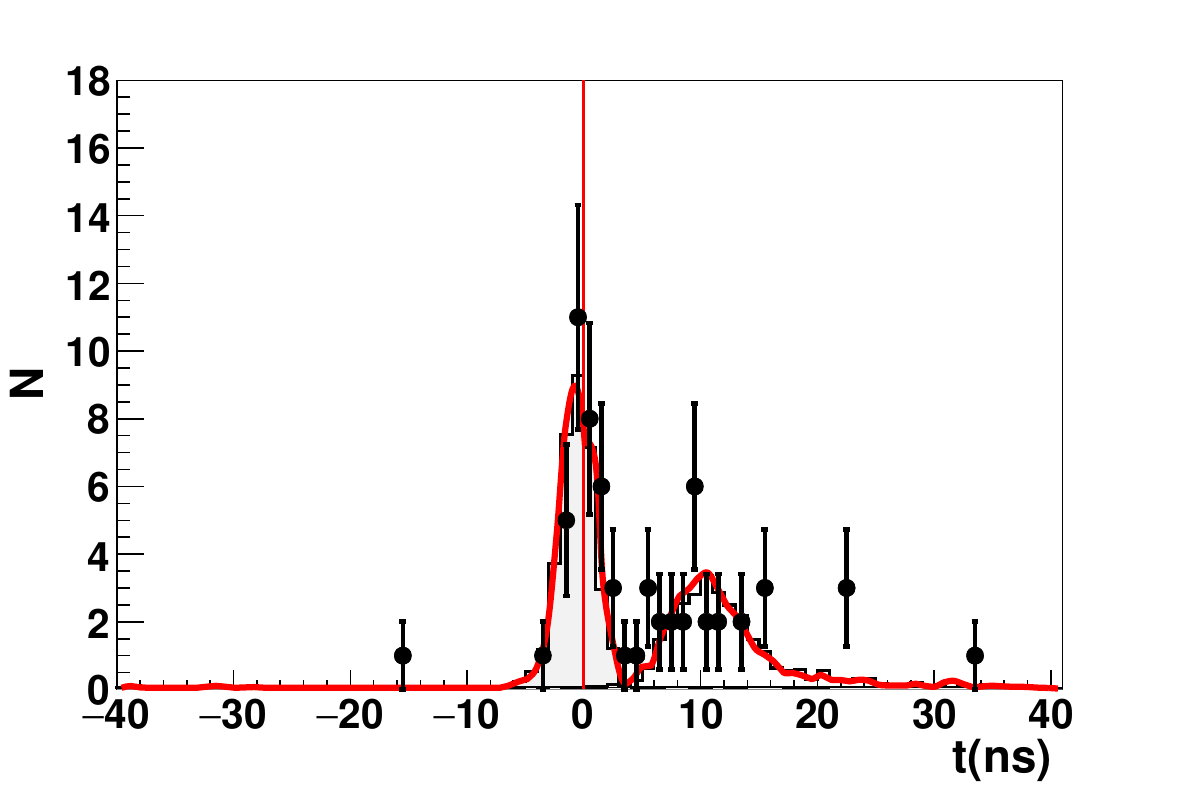} \hfill
\includegraphics[width=0.48\textwidth]{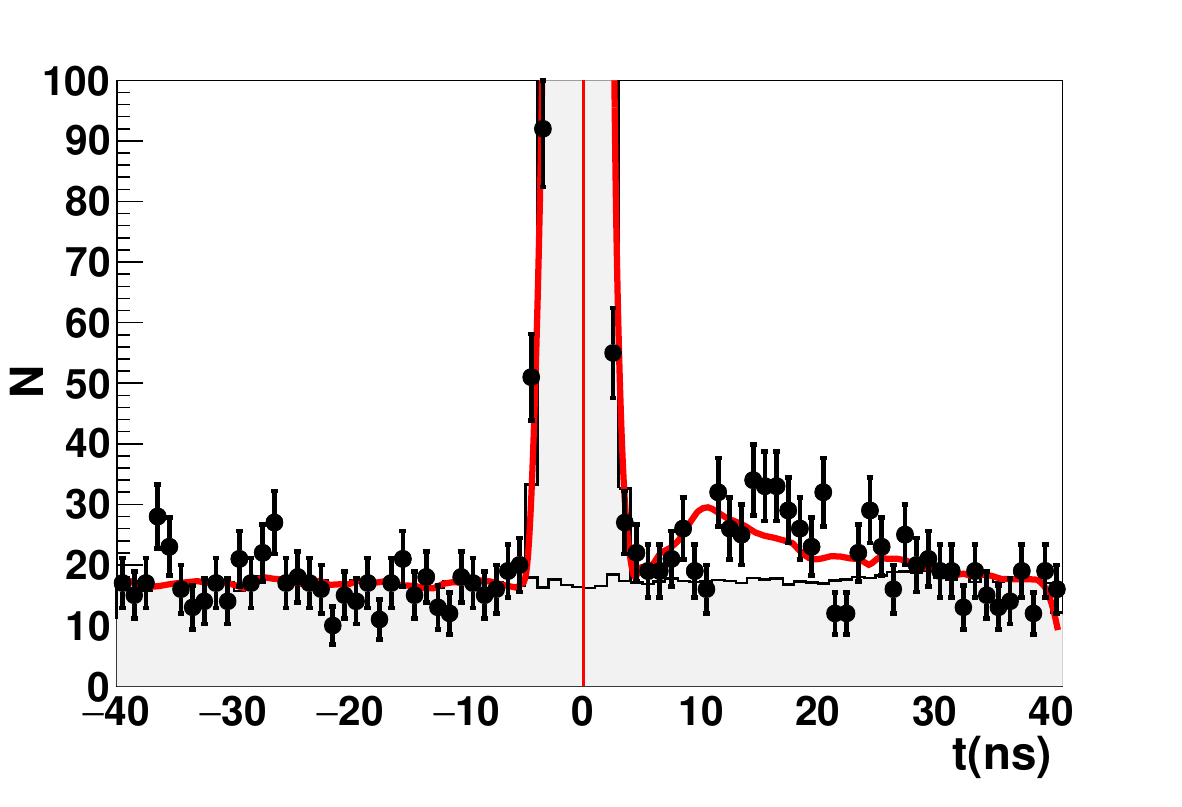} \\
\parbox[h]{0.48\textwidth}  {\caption { The event time spectrum with 
tracks in drift chamber at $E_b$=943.5 MeV. The peak at 
t=0 corresponds to the beam and physical background. The wide peak 
from the right is the $n\bar{n}$ signal. }
\label{fig:nchcor}} \hfill
\parbox[h]{0.48\textwidth} {\caption{ The event time spectrum with inverted 
EMC energy cut $E_{EMC}<E_b$. The peak at t=0 corresponds to the beam
and physical background. The wide peak from the right  is the delayed   
$n\bar{n}$  signal. }
\label{fig:etcor}}
\end{figure*}

The detection efficiency in our measurement is of order of  20\%,
what is insufficient  for  such a complex process to analyze  as 
$e^+e^-\to n\bar{n}$.  In the list of selection conditions 
(chapter \ref{sec:EvSelect}), the greatest contribution to the loss of 
efficiency is made by the cosmic background suppression parameter 
shcosm (number 6 in the list), where the loss of  efficiency 
is about 50\%. All other selection conditions result in a
significantly smaller efficiency loss $\sim$ 1-20\% for each. For
correct calculation of efficiency in general,  
it is important to find out how correctly the proportion of events
outside the selection condition is simulated. 
To do this, we invert the selection conditions for each 
selection condition or a pair of related conditions and then
calculate the corresponding corrections $\delta$ for detection
efficiency in each of 13 energy points as follows:
\begin{equation}
\delta=\frac{n_0}{n_0+n_1}\frac{m_0+m_1}{m_0},
\label{effcor}
\end{equation}
where $n_0$ ($n_1$) is the number of $n\bar{n}$ data events determined 
with standard (inverted) selection cuts. These numbers were calculated 
during the time spectra fitting with the Eq.\ref{eq17}, as it is described in
the chapter \ref{sec:Tim19}.  The values $m_0$ and $m_1$ refer respectively 
to the MC simulation event numbers.   Examples of the time spectra 
obtained with inverted conditions are shown in
Figs.~\ref{fig:nchcor},~\ref{fig:etcor}.

  Below we will give some comments on obtaining correction factors
$\delta$ for   detection efficiency. During the initial selection of
events for $n\bar{n}$ analysis, events with tracks not from the
interaction region are allowed. The condition for such tracks is
$D_{xy}>0.5$ cm, where $D_{xy}$ is the distance between the track and 
the axis of the beams. Accordingly, the events with $D_{xy}<0.5$ are
rejected. There is no significant loss of $n\bar{n}$ events due to
this conditions. On the other side,  the events of the beam and physical 
background, with tracks predominantly from the beam region, 
are strongly suppressed. When inverting the selection condition on the
number of charged tracks (nch$>$0), it is worth to consider the possible
background  from the  related process $e^+e^-\to p\bar{p}$. 
In this experment at the beam energy  $E_b < 955$ MeV the protons
and antiproptons are slow and stop in the collider vacuum pipe. 
Here the antiproton annihilates  with the production  of charged
tracks, most of which have $D_{xy}>0.5$ cm and are included in the
inverse selection nch$>$0. 
But even such a background from the $e^+e^-\to
p\bar{p}$ will be suppressed when fitting the time spectrum, since it
practically merges in time with the beam background. An example of a
time spectrum with inverse selection condition nch$>$0 is shown in 
Fig.~\ref{fig:nchcor}. 

\begin{figure*}
\includegraphics[width=0.48\textwidth]{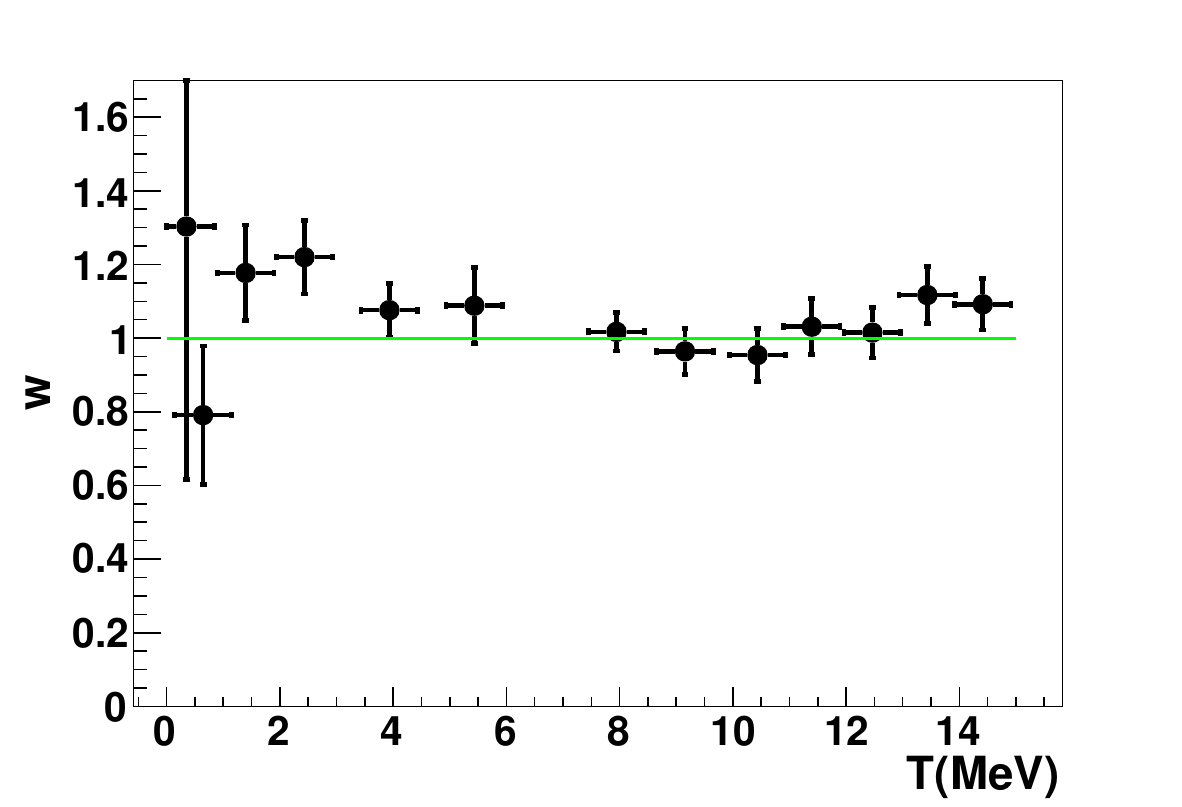} \hfill
\includegraphics[width=0.48\textwidth]{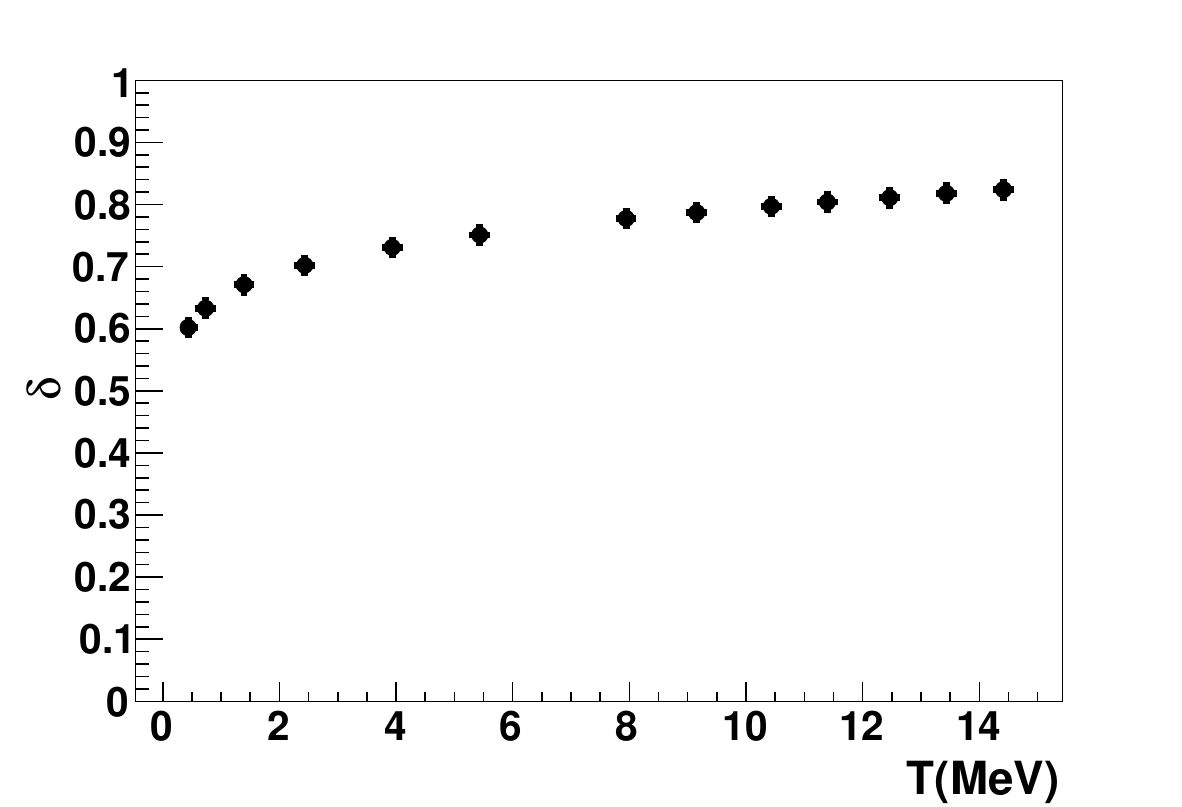} \\
\parbox[h]{0.48\textwidth}  {\caption { The total detection efficiency
correction, obtained using Eq.\ref{eqcorr}, versus antineutron kinetic energy 
$T=E_b-m_n$. Horizontal line corresponds to no correction case. }
\label{fig:epscors}} \hfill
\parbox[h]{0.48\textwidth} {\caption{ The radiative correction
to the  $e^+e^-\to n\bar{}$ process as a function of kinetic 
energy $T=E_b-m_n$.}
\label{fig:rcor}}
\end{figure*}

To study the effect of the energy threshold in the EMC, the inverse
condition  $0.7 E_b<E_{\rm cal}<E_b$ is applied. An example of time
spectrum under this condition is shown in Fig.~\ref{fig:etcor}. Here,
in contrast to  Fig.~\ref{fig:nchcor}, a sharply increased beam and
cosmic background is seen. However, the $n\bar{n}$ signal is
clearly visible here too, which makes it possible to calculate the
efficiciency correction. 

An additional correction arises from the events with EMC energy 
$E_{\rm cal}<0.7E_b$. These events are not recorded in the Ntuples for 
analysis due to the large baqckground. In events with such low EMC energy
antineutron is absorbed usualy in the 3-d calorimeter layer or outside  
the calorimeter.
Fortunately, the proportion of such events is small: 3.5\% at $E_b$=954
MeV and 1\% at  $E_b$=940 MeV. Monte Carlo studies show that these
events are solely represented by the contribution of the antineutron
scattering process. It was previously noted in chapter ~\ref{sec:Tim19},  
that to describe the shape of data time spectrum
the contribution of the process of antineutron scattering in MC
should be reduced by a factor of 1.5-2. With such a change, the
proportion of events with  $E_{\rm cal}<0.7 E_b$ in MC reduces to 0.7\% at
$E_b$=940 MeV and to 2\% at  $E_b$=954 MeV. The difference between the
initial and modified values is taken as an error to the final
correction. The sign of this correction is positive,  it icreases the
detection efficiency by several percent. 

Next, we calculated the corrections to the detection efficiency for
all selection conditions (chapter \ref{sec:EvSelect}) using
Eq.\ref{effcor}. Then, considering the corrections to be uncorrelated,
we calculate the total correction for a given energy as:
\begin{equation}
\delta_{\rm tot}=\Pi(1+\delta_i)-1.
\label{eqcorr}
\end{equation}
Its energy dependence shown in Fig.~\ref{fig:epscors}. It can be seen,
that if we exclude two energy points near the threshold, the
total correction increases the   $e^+e^-\to n\bar{n}$ process
detection efficiency by $\sim$10\%.
   The corrected detection efficiency is obtained from the MC efficiency
by multiplying by the total correction $\delta_{tot}$.    
The values of the corrected efficiency are given along with systematic 
errors in the Table~\ref{tab:crsect} and shown in Fig.~\ref{fig:efenr}.

\section{The measured $e^+e^-\to n\bar{n}$ cross section \label{sec:Crosct}}
\begin{figure*}[h]
\includegraphics[width=0.49\textwidth]{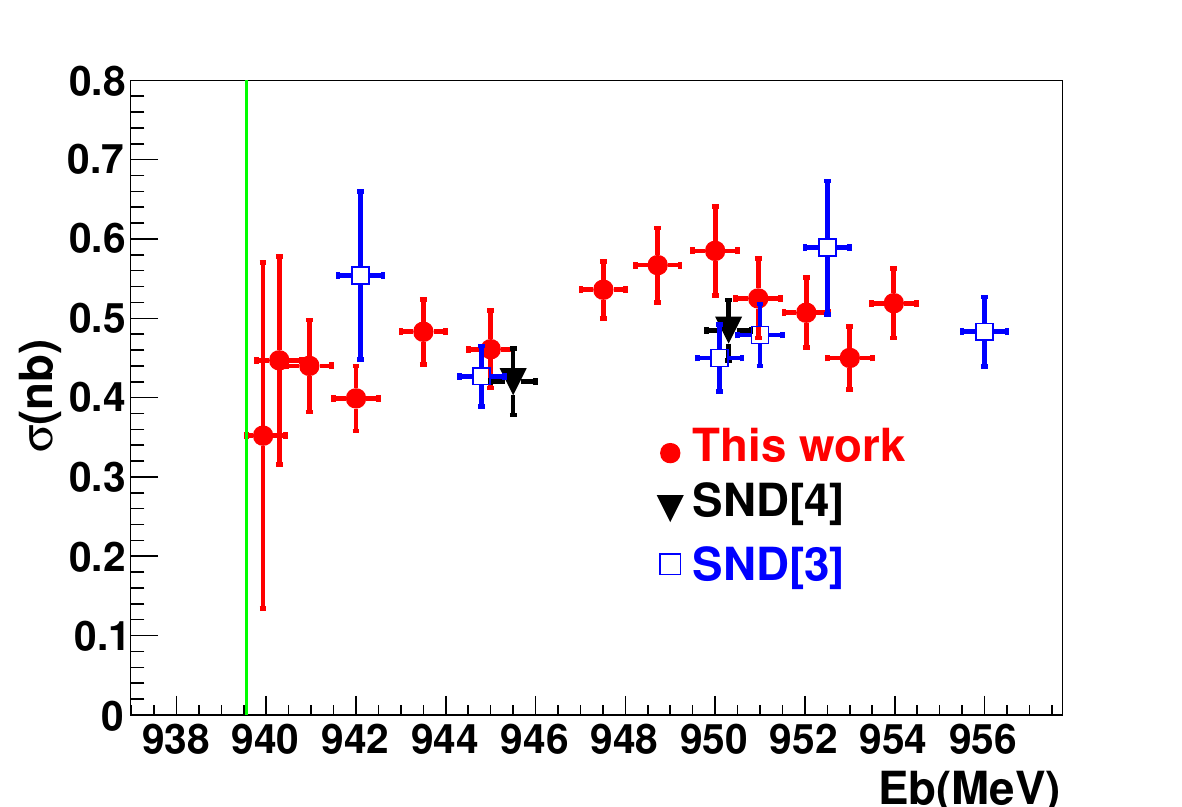} \hfill
\includegraphics[width=0.49\textwidth]{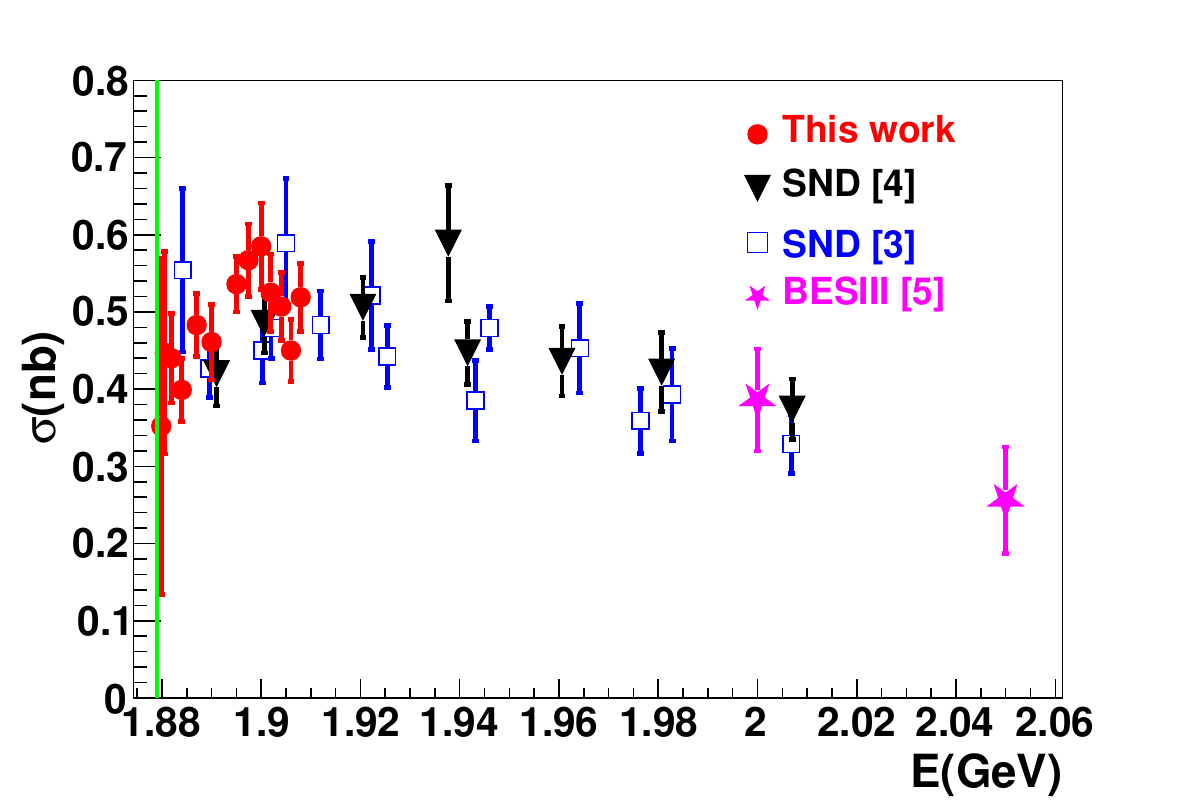} \\
\parbox[h]{0.45\textwidth}{
\caption {The measured $e^+e^-\to n\bar{n}$ cross section in the
vicinity of the nucleon threshold. The vertical line is the $n\bar{n}$
threshold.  }
\label{fig:csnnE1}} \hfill
\parbox[h]{0.45\textwidth}{
\caption { The measured $e^+e^-\to n\bar{n}$ cross section as a function of
neutron momentum  in comparison with previous measurements. 
The vertical line is the $n\bar{n}$ threshold. }
\label{fig:csnnE2}}
\end{figure*}
Using the number of $n\bar{n}$ events $N_{n\bar{n}}$,
luminosity $L$ and detection
efficiency $\varepsilon$ (Table \ref{tab:crsect}), the
visible cross section  $\sigma_{vis}(E)= N_{n\bar{n}}/L\varepsilon$   
can be calculated. The Born cross section $\sigma(E)$, that we need,  is
related to the visible cross  $\sigma_{vis}(E)$ in the following form :
\begin{eqnarray}
\sigma_{vis}(E)&=&\sigma(E)(1+\delta(E))\nonumber\\
&=&\int_{-\infty}^{+\infty}G(E^\prime,E)dE^\prime\nonumber\\
&&\int_0^{x_{max}}W(s,x)\sigma(s(1-x))dx,
\label{eqB4}
\end{eqnarray}
where  $W(s,x)$ is the radiator function~\cite{Radcor},  describing
emission of  photons with energy $xE_b$ by initial electrons and positrons, 
$G(E^\prime,E)$ is a Gaussian function describing the c.m. energy spread.
In  function $W(s,x)$ the 
contribution of the vacuum polarization is not taken  into account,  
so our Born cross section is the ``dressed'' cross section.
The factor $(1+\delta(E))$ takes into account both the radiative
corrections and beam energy spread. 
This factor (Fig.\ref{fig:rcor}) is calculated in each of 13 energy points 
using the Born cross
section, obtained by the fitting of the visible cross section using 
Eq.~\ref{eqB4}. The energy dependence of the Born cross section   
is described by Eq.\ref{eqB2}, 
in which the form factor is a 2-nd order polynomial of
the neutron momentum. The parameters of this polynomial were free
fit  parameters.
The resulting  Born cross section is shown in the Figs.~\ref{fig:csnnE1},
~\ref{fig:csnnE2} as a function of the neutron energy and momentum.
Numerical values of the cross
section are given in the Table~\ref{tab:crsect}. The dominant contribution 
into systematic error in the measured cross section 
is made by the detection efficiency correction error,
In this error the uncertainties in the value of luminosity (2\%) and 
radiative correction (2\%) are also taken into account. 
   In Figs.~\ref{fig:csnnE1},~\ref{fig:csnnE2}   the total
statistical and systematic error is shown. 
In comparison with previous
SND measurements ~\cite{Art1719}, \cite{Art2021}, this paper presents data  
in the immediate vicinity of the $n\bar{n}$ threshold. 
At the beam energy points of 942, 945 and 950 MeV, the
obtained cross section agrees with  the previous SND results
At energy points extremely close to the threshold
(939.6, 940.2 MeV) the measured cross section does not drop to zero and is
approximately 2 times lower than the  $e^+e^-\to p\bar{p}$ cross section
\cite{Babar},\cite{Babar1}.

\section{The neutron effective timelike formfactor \label{sec:fform} }
\begin{figure*}[h]
\includegraphics[width=0.49\textwidth]{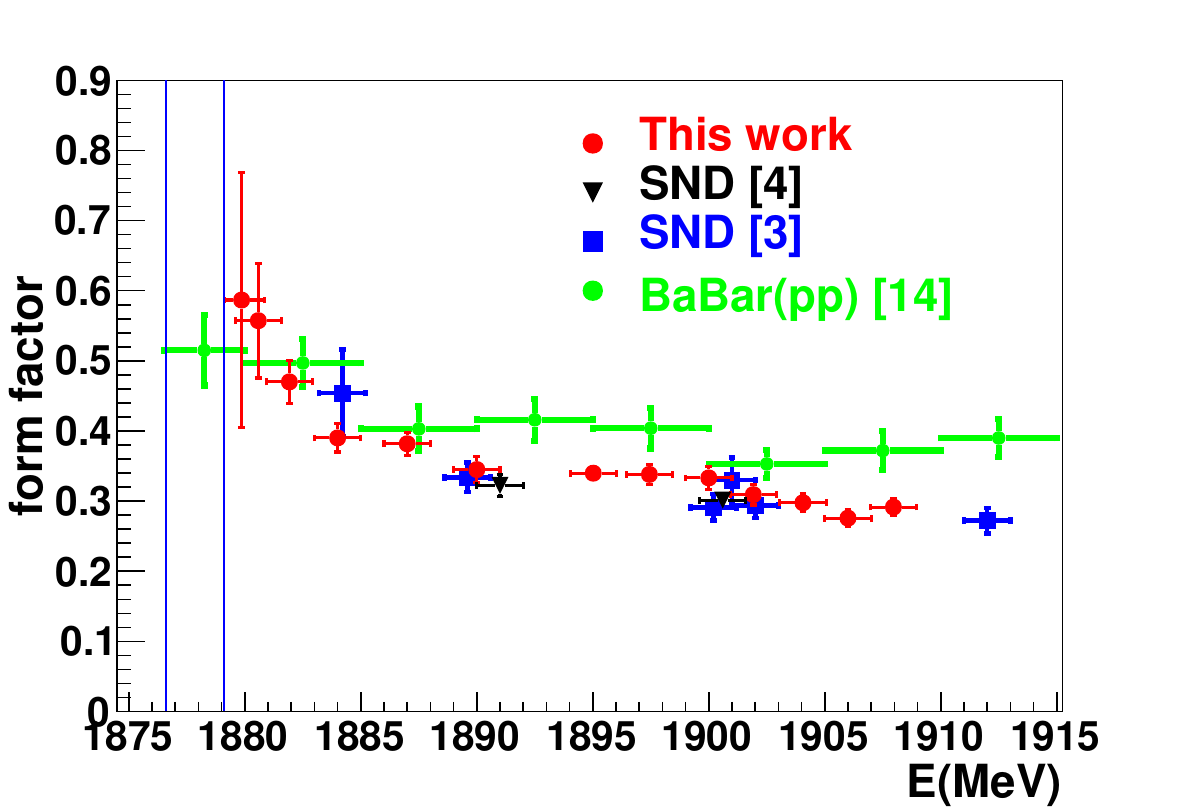} \hfill
\includegraphics[width=0.49\textwidth]{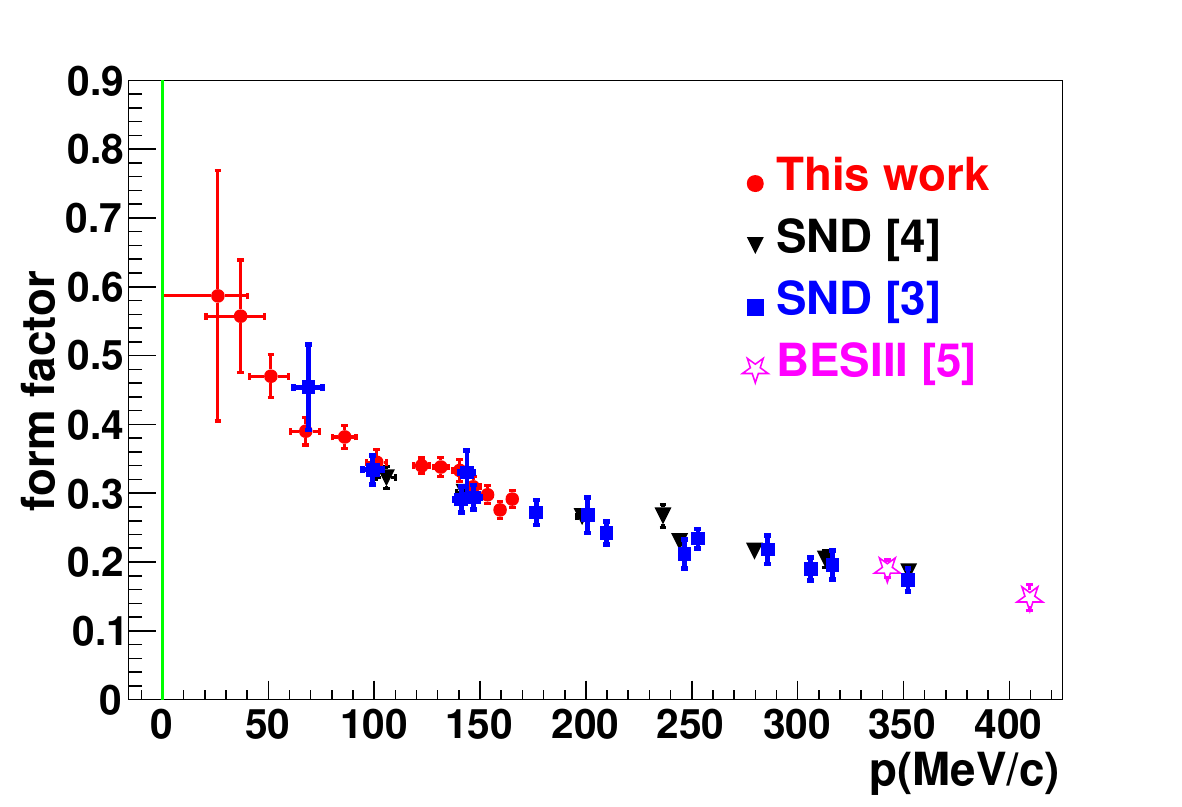} \\
\parbox[h]{0.45\textwidth}{
\caption {The measured effective neutron timelike form factor
in the vicinity of the nucleon threshold. For comparison the  
proton form factor measured in the Babar experiment \cite{Babar} is
shown. Vertical lines correspond to the position of the  proton and neeutronthreshold. }
\label{fig:ffnnE}} \hfill
\parbox[h]{0.45\textwidth}{
\caption {
The measured effective neutron timelike form factor
as a function of neutron momentum,
compared with previous  measurements. }
\label{fig:ffnnP}}
\end{figure*}

The effective neutron form factor calculated from the measured cross 
section using Eq.~(\ref{eqB2}) is listed in the Table \ref{tab:crsect}
and shown in Fig.~\ref{fig:ffnnE} as a function of the antineutron
energy 
and in Fig.~\ref{fig:ffnnP} as a function of the antineutron momentum.
The results of previous measurements are also shown. For comparison
the Fig.~\ref{fig:ffnnE} shows the proton form factor measured in the
Babar experiment \cite{Babar},\cite{Babar1}.  It is seen that closer 
to the threshold the neutron formfactor value is  about 0.6 and  
the form factors of the neutron and proton approach each other, 
although the error in the measurement of neutrons is still large $\sim$30\%.
Most predictions of nucleon form factors  values refer to the asymptotic region
of large momentum transfers, there  is no unambiguous prediction for
our case of the threshold region.

\section{Summary \label{sec:summary} }
The experiment to measure the $e^+e^-\to n\bar{n}$ cross section 
and the neutron timelike form factor has been
carried out with the SND detector at the VEPP-2000 $e^+e^-$ collider
at energies in c.m. from the  process threshold to 1908 MeV.
The measured cross section varies  with energy  within 0.4$\div$0.6 nb. 
At the energy point closest to the threshold, the cross section is
about 0.4 nb. There are no contradictions with previous SND
measurements. The  neutron effective timelike form factor, extracted from the
measured cross section,  shows an
increase to a value of 0.6 at the very threshold. A tendency is
visible for the form factors of the neutron and proton to converge at
the threshold.

{\bf ACKNOWLEDGEMENTS}. This work was carried out on the Russian
Science Foundation   grant No. 23-22-00011.

\end{document}